\documentclass[aps, prl, letterpaper, amsmath, amssymb, preprintnumbers, showpacs, superscriptaddress, twocolumn, floatfix, longbibliography, nofootinbib]{revtex4-1}
\usepackage{xspace}
\usepackage{graphicx}
\usepackage[dvipsnames]{xcolor}
\pdfoutput=1 

\allowdisplaybreaks 

\usepackage[hypertexnames=false]{hyperref}
\hypersetup{
    colorlinks=true,       
    linkcolor=blue,          
    citecolor=blue,        
    filecolor=blue,      
    urlcolor=blue           
}
\usepackage{slashed}
\usepackage{mathtools}

\newcommand{\nnbar}{\ensuremath{n\text{-}\overline{n}}}

\long\def\/*#1*/{}
\definecolor{red}{rgb}{1.0, 0, 0}

\newcommand{\ket}[1]{\left| #1 \right>} 
\newcommand{\mbraket}[3]{\left< #1 \vphantom{#2#3} \right|
 #2 \left| #3 \vphantom{#1#2} \right>} 

\newcommand{\lbsm}{\ensuremath{\Lambda_\text{BSM}}}
\newcommand{\lqcd}{\ensuremath{\Lambda_\text{QCD}}}
\newcommand{\msbar}{\ensuremath{\overline{\text{MS}}}}

\begin{document}

\title{Neutron-antineutron oscillations from lattice QCD}

\author{Enrico Rinaldi}
\email{erinaldi@bnl.gov}
\affiliation{RIKEN BNL Research Center, Brookhaven National Laboratory, Upton, NY 11973, USA}
\affiliation{Nuclear Science Division, Lawrence Berkeley National Laboratory, Berkeley, CA 94720, USA}
\author{Sergey Syritsyn}
\email{sergey.syritsyn@stonybrook.edu}
\affiliation{RIKEN BNL Research Center, Brookhaven National Laboratory, Upton, NY 11973, USA}
\affiliation{Department of Physics and Astronomy, Stony Brook University, Stony Brook, NY 11794, USA}
\author{Michael L. Wagman}
\email{mlwagman@mit.edu}
\affiliation{Center for Theoretical Physics, Massachusetts Institute of Technology, Cambridge, MA 02139, USA}
\author{Michael I. Buchoff}
\affiliation{Lawrence Livermore National Laboratory, Livermore, California 94550, USA}
\author{Chris Schroeder}
\affiliation{Lawrence Livermore National Laboratory, Livermore, California 94550, USA}
\author{Joseph Wasem}
\affiliation{Lawrence Livermore National Laboratory, Livermore, California 94550, USA}

\begin{abstract}

Fundamental symmetry tests of baryon number violation in low-energy experiments can probe beyond the
Standard Model (BSM) explanations of the matter-antimatter asymmetry of the universe.
Neutron-antineutron oscillations are predicted to be a signature of many baryogenesis mechanisms
involving low-scale baryon number violation.
This work presents first-principles calculations of neutron-antineutron matrix elements
needed to accurately connect measurements of the neutron-antineutron oscillation
rate to constraints on $|\Delta B|=2$ baryon number violation in BSM theories.
Several important systematic uncertainties are controlled by using a state-of-the-art lattice gauge field ensemble with physical quark masses and approximate chiral symmetry,
performing nonperturbative renormalization with perturbative matching to the $\msbar$ scheme, and studying excited state effects in two-state fits.
Phenomenological implications are highlighted by comparing expected bounds from proposed
neutron-antineutron oscillation experiments to predictions of a specific model of post-sphaleron
baryogenesis.
Quantum chromodynamics is found to predict at least an order of magnitude more events in neutron-antineutron oscillation experiments
than previous estimates based on the ``MIT bag model'' for fixed BSM parameters.
Lattice artifacts and other systematic uncertainties that are not controlled in this pioneering
calculation are not expected to significantly change this conclusion.

\end{abstract}
\preprint{RBRC-1290, MIT-CTP/5051, LLNL-JRNL-757017}

\maketitle

\textbf{\textit{Introduction --- }}\label{sec:introduction}
Beyond the Standard Model (BSM) violation of baryon number conservation is necessary to explain the
observed matter-antimatter asymmetry of the universe.
Baryogenesis explanations involving physics at high scales, such as leptogenesis, are appealing but
difficult to test in low-energy experiments.
Alternative explanations such as post-sphaleron baryogenesis involve low-scale baryon number
violation that can be directly constrained by current and near-future experiments.
Extracting robust BSM theory constraints from these experiments is critical for fundamental symmetry
tests of baryon and lepton number violation addressing the long-standing mystery of
matter-antimatter asymmetry.

Neutron-antineutron oscillations ($\nnbar$) are predicted as a signature of low-scale baryogenesis
in BSM theories including $SO(10)$ grand unified theories (GUTs), left-right symmetric theories,
$R$-parity-violating supersymmetry, low-scale quantum gravity, extra-dimensional theories, and
string theories with exotic instantons; see Refs.~\cite{Mohapatra2009,Babu2013,Phillips2016} for
comprehensive reviews and further references.
Experimental constraints on $\nnbar$ come from large underground detection experiments such as
Super-K~\cite{Abe2015} and SNO~\cite{Aharmim2017} or from cold neutron time-of-flight
experiments~\cite{Baldo-Ceolin1994}.
Results are presented as bounds on the neutron-antineutron oscillation time $\tau_{\nnbar}$ governing the
time-dependent probability $P_{\nnbar} = \sin^2(t/\tau_{\nnbar})$ for a free neutron in vacuum to
turn into an antineutron~\cite{Phillips2016}.

The best direct bound on $\tau_{\nnbar}$ is from the cold neutron experiment at Institut
Laue-Langevin (ILL), $\tau_{\nnbar} > 0.89 \times 10^8 \ \text{s}$~\cite{Baldo-Ceolin1994}, which is
essentially background-free and can be improved with larger neutron flux, magnetic
shielding, and the latest technologies in neutron optics.
There has been a proposal for a new cold neutron experiment at the European Spallation Source (ESS)
that would be $\approx10^3$ times more sensitive than the ILL experiment and improve cold neutron
constraints on $\tau_{\nnbar}$ by a factor of 32~\cite{Milstead2015,Frost2016},
as well as experiments at other reactors~\cite{Fomin2018}.
The best bound to date on the neutron-antineutron transition time within
Oxygen-16 is from Super-K, $\tau_{O^{16}} > 1.9 \times 10^{32} \ \text{years}$~\cite{Abe2015}, which
constrains the free oscillation time $\tau_{\nnbar} > 2.7 \times 10^8 \ \text{s}$ after taking into
account nuclear structure effects~\cite{Friedman2008}.
In the future, we expect underground neutrino facilities like DUNE to be able to provide competitive
bounds thanks to improved background-rejection techniques~\cite{Hewes2017}.

BSM theories cannot directly predict $\tau_{\nnbar}$, which also depends on low-energy dynamics of
quantum chromodynamics (QCD) responsible for binding quarks into hadrons.
Since BSM and QCD effects are important at different scales, any BSM mechanism mediating $\nnbar$
oscillations can be summarized into a particular linear combination of effective six-quark operators
violating baryon number by two units.
Their matrix elements between neutron and antineutron states are determined by nonperturbative QCD,
and have to be computed before any BSM predictions for $\tau_{\nnbar}$ can be made.
Prior to this work, only ``MIT bag model'' estimates of these six-quark matrix
elements~\cite{Rao1982,Rao1984} have been available
and connections between $\tau_{\nnbar}$ and new physics constraints have included unknown model uncertainties.

The purpose of this letter is to present a lattice QCD (LQCD) calculation of these six-quark matrix
elements and to discuss its impact on possible discovery of new physics and theories of
baryogenesis.
Our work improves upon preliminary LQCD results~\cite{Buchoff2012} by using chiral quarks at the
physical point and robust nonperturbative renormalization, and upon recent
results~\cite{Syritsyn2015} by analyzing excited state effects.
With most of LQCD uncertainties under control, we find that QCD predicts 1 to 2
orders of magnitude larger rate of $\nnbar$ oscillations than previously expected.
Our results have recently been presented~\cite{enrico_rinaldi_2019_2539735} at workshops dedicated
to devising a plan to observe $\nnbar$ oscillations at the ESS, future reactors and underground
laboratories~\cite{ess_workshop_2018}, and used in generic effective field theories (EFTs)
for baryogenesis such as the recent work in Ref.~\cite{Grojean2018}.

\textbf{\textit{Neutron-antineutron operators --- }}\label{sec:operators}
The basis for the lowest-dimension operators for $\nnbar$ transitions that are color singlets and
electrically neutral was constructed in
Refs.~\cite{Chang:1980ey,Kuo1980,Rao1982,Rao1984,Caswell:1982qs}
\begin{equation}
\label{eqn:op_nnbar_orig}
  \begin{split}
    \mathcal{O}^1_{\chi_1 \chi_2 \chi_3} &= (u^T CP_{\chi_1} u) (d^T CP_{\chi_2} d) (d^T CP_{\chi_3} d)T^{SSS}\,, \\
    \mathcal{O}^2_{\chi_1 \chi_2 \chi_3} &= (u^T CP_{\chi_1} d) (u^T CP_{\chi_2} d) (d^T CP_{\chi_3} d)T^{SSS}\,, \\
    \mathcal{O}^3_{\chi_1 \chi_2 \chi_3} &= (u^T CP_{\chi_1} d) (u^T CP_{\chi_2} d) (d^T CP_{\chi_3} d)T^{AAS}\,,
  \end{split}
\end{equation}
where quark spin indices are implicitly contracted in the parentheses and quark color indices are
implicitly contracted with the tensors
\begin{equation}
  \nonumber
  \begin{split}
	T^{SSS}_{\{ij\}\{kl\}\{mn\}} &= \varepsilon_{ikm}\varepsilon_{jln} + \varepsilon_{jkm}\varepsilon_{iln}
        + \varepsilon_{ilm}\varepsilon_{jkn} + \varepsilon_{ikn}\varepsilon_{jlm}\,,\\
        T^{AAS}_{[ij][kl]\{mn\}} &= \varepsilon_{ijm}\varepsilon_{kln} +
        \varepsilon_{ijn}\varepsilon_{klm}\,,
  \end{split}
\end{equation}
$P_{L,R} = \frac{1}{2}(1 \mp \gamma_5)$ are chiral projectors, and the Euclidean charge conjugation
matrix $C$ satisfies $C\gamma_\mu C^\dag = -\gamma_\mu^T$.
Symmetries together with the Fierz relation $ \mathcal{O}^2_{\chi\chi\chi^\prime} -
\mathcal{O}^1_{\chi\chi\chi^\prime} = 3\mathcal{O}^3_{\chi\chi\chi^\prime}$ reduce the number of
independent operators to 14 (7+7 related by parity).
In this work, we use basis operators from chiral isospin multiplets that renormalize
multiplicatively~\cite{Syritsyn2015,Buchoff2016} and have particular implications for
phenomenology.
An extended discussion of the operators and their chiral properties is reported in Ref.~\cite{Rinaldi:2019thf}.
There are three Standard Model (SM) gauge-singlet operators\footnote{
  Two additional operators arise in dimensional regularization at two-loop order~\cite{Buchoff2016}
  that are exactly equal to $Q_{1,3}$ in lattice regularization by Fierz identities. These Fierz
  identities are preserved in the modified minimal subtraction ($\msbar$) scheme if one-loop matching is consistently included with two-loop
  running. At high scales $\lbsm \gg \lqcd$ Fierz identity violations can be neglected even if BSM
  matching is performed at tree level.}
that provide the dominant contributions to $\nnbar$ transitions in SM EFT,
\begin{equation}
\label{eqn:op_nnbar_123}
  \begin{split}
	Q_1 &= -4(u^T CP_R d)(u^T CP_R d)(d^T CP_R d)T^{AAS},\\
	Q_2 &= -4(u^T CP_L d)(u^T CP_R d)(d^T CP_R d)T^{AAS},\\
	Q_3 &= -4(u^T CP_L d)(u^T CP_L d)(d^T CP_R d)T^{AAS}.
\end{split}
\end{equation}
The fourth $SU(2)_L\times U(1)$ singlet operator
\begin{equation}
\label{eqn:op_nnbar_4}
  \begin{split}
    Q_4 &= - \frac{4}{5}(u^T CP_R u)(d^T CP_R d)(d^T CP_R d)T^{SSS} \\
    &\hspace{15pt} -\frac{16}{5}(u^T CP_R d)(u^T CP_R d)(d^T CP_R d)T^{SSS}
\end{split}
\end{equation}
has vanishing matrix elements in the $SU(2)$ isospin limit and is not studied in this work.
Isospin-violating corrections provide a subdominant systematic uncertainty and are also neglected.
$SU(2)_L$-non-singlet operators can also lead to $\nnbar$ oscillations,
\begin{equation}
\label{eqn:op_nnbar_567}
  \begin{split}
        Q_5 &= (u^T CP_R u)(d^T CP_L d)(d^T CP_L d)T^{SSS},\\
	Q_6 &= -4(u^T CP_R d)(u^T CP_L d)(d^T CP_L d)T^{SSS},\\
        Q_7 &= - \frac{4}{3}(u^T CP_L u)(d^T CP_L d)(d^T CP_R d)T^{SSS} \\
	&\hspace{15pt} -\frac{8}{3}(u^T CP_L d)(u^T CP_L d)(d^T CP_R d)T^{SSS} .
  \end{split}
\end{equation}
These operators transform in the same chiral irreducible representation but describe different
chiral multiplet components and do not mix under renormalization~\cite{Buchoff2016,Rinaldi:2019thf}.
Isospin $SU(2)$ symmetry leads to the following relation between matrix elements
\begin{equation}
  \begin{split}
    \mbraket{\overline{n}}{Q_5}{n} &= \mbraket{\overline{n}}{Q_6}{n} = -\frac{3}{2}\mbraket{\overline{n}}{Q_7}{n}
  \end{split}
\label{eqn:op_nnbar_chiral567}\,.
\end{equation}

The complete chiral basis of QCD and QED singlet operators is given by $Q_I,\,Q_I^P$, $I=1\ldots7$
where parity-transformed operators are $Q_I^P=(-Q_I)|_{L\leftrightarrow R}$.
The operators~(\ref{eqn:op_nnbar_123}-\ref{eqn:op_nnbar_567}) are related to the ones in
Eq.~(\ref{eqn:op_nnbar_orig}) as
\begin{equation}
  \begin{split}
    Q_1 &= -4\mathcal{O}^3_{RRR}, \hspace{5pt} Q_2
      = -4\mathcal{O}^3_{LRR}, \hspace{5pt} Q_3 = -4\mathcal{O}^3_{LLR}, \\
    Q_4 &= -\frac{4}{5}\mathcal{O}^1_{RRR} - \frac{16}{5}\mathcal{O}^2_{RRR}, \hspace{10pt} Q_5
      = \mathcal{O}^1_{RLL}, \\
    Q_6 &= -4 \mathcal{O}^2_{RLL}, \hspace{10pt} Q_7
      = -\frac{4}{3}\mathcal{O}^1_{LLR} - \frac{8}{3}\mathcal{O}^2_{LLR}.
  \end{split}\label{eq:basisdef}
\end{equation}
Because of symmetries and Eq.~\eqref{eqn:op_nnbar_chiral567}, only four
separate nucleon matrix elements $\langle \bar{n} | Q_{1,2,3,5}|n\rangle$ need to be determined
using lattice QCD methods.

\begin{table}[t!]
\begin{tabular}{||c||r|r|r|r||}
   \hline
   \text{Operator}
  & $\mathcal{M}_I^{\msbar}$
  & $\mathcal{M}_I^{\msbar}$
  & $\frac{ \mathcal{M}_I^{\msbar}}{\text{MIT bag A}}$
  & $\frac{ \mathcal{M}_I^{\msbar}}{\text{MIT bag B}}$ \\
  & (2 GeV)
  & (700 TeV)
  & (2 GeV)
  & (2 GeV) \\
    \hline
   $Q_1$ &  $-46(13)$ & $-26(7)$ & 4.2  & 5.2    \\\hline
   $Q_2$ &  $95(17)$ & $144(26)$ & 7.5 &  8.7    \\\hline
   $Q_3$ &  $-50(12)$ &  $-47(11)$ & 5.1 & 6.1    \\\hline
   $Q_5$ &  $-1.06(48)$ & $-0.23(10)$ & -0.8 & 1.6    \\\hline
  \end{tabular}
  \caption{For each operator $Q_I$ we
    show its renormalized matrix element value $\mathcal{M}_I^{\msbar}$ in units of
    [$10^{-5} \text{ GeV}^6$].
    The total uncertainty includes statistical and systematic errors added in quadrature.
    Renormalized results are obtained through nonperturbative RI-MOM renormalization and
    perturbative matching to $\msbar$ at two scales: 2 GeV and 700 TeV in column two and three,
    respectively.
    The last two columns show a comparison between the Lattice QCD matrix elements and the
    results of the same matrix elements for two choices of the ``MIT bag model'' from Ref.~\cite{Rao1982}.
 \label{tab:renorm_me}}
\end{table}

\textbf{\textit{Lattice QCD results --- }}\label{sec:lat_calc}
For this calculation, we use an ensemble of QCD gauge field configurations on a $48^3\times96$ lattice
generated with Iwasaki gauge action and $N_f=2+1$ flavors of dynamical M\"obius Domain Wall fermions
with masses almost exactly at the physical point~\cite{Blum:2014tka}.
The pion mass is $m_\pi=139.2(4)\text{ MeV}$ and the lattice spacing is $a=0.1141(3)\text{ fm}$.
With the physical lattice size $L\approx5.45\text{ fm}$, and $m_\pi L = 3.86$, finite volume effects
on the matrix elements are estimated from chiral perturbation theory to be $\lesssim
1\%$~\cite{Bijnens2017}.
Discretization effects are expected to be small because the meson decay constants
$f_{\pi,K}$~\cite{Blum:2014tka}, the nucleon mass and dispersion~\cite{Syritsyn:2019vvt} on this
ensemble are very close to physical.

We calculate lattice correlation functions on 30 independent gauge field configurations separated by
40 MD steps using point-source quark propagators aided by all-mode-averaging (AMA) sampling~\cite{Shintani:2014vja} to
reduce stochastic uncertainty.
On each configuration, we compute 1 exact and 81 low-precision samples evenly distributed over the
4D volume.
For the latter, quark propagators are computed with low-mode deflation and 250 iterations of the
conjugate gradient algorithm.
The propagators are contracted into intermediate baryon blocks~\cite{Doi:2012xd,Detmold:2012eu}
representing (anti)neutron source or sink operators made of point and Gaussian-smeared quarks and
denoted by $n^{J=P,S}$, respectively.
These blocks are finally contracted into (anti)neutron two-point correlation functions of $P$ source
and $J=P,S$ sink operators,
\begin{equation}
G_{2pt}^{PJ}(t)
  = \sum_{\mathbf{x}} \langle n_\uparrow^J(\mathbf{x},t) \overline{n}_\uparrow^P(0) \rangle
  = \sum_{\mathfrak{n}} \sqrt{ Z_\mathfrak{n}^{J} Z_\mathfrak{n}^{P} } e^{-E_\mathfrak{n} t} \,,
\label{eq:2ptspec}
\end{equation}
as well as three-point correlation functions
involving $J=S,P$ antineutron sources, neutron sinks, and six-antiquark operators $\bar{Q}_I$
that are obtained from $Q_I$ by charge conjugation and have identical matrix elements,
\begin{align}
\nonumber
&G_{3pt}^{JJ^\prime}(\tau,t;Q_I)
  = \sum_{\mathbf{x},\mathbf{y}} \langle n_\uparrow^J(\mathbf{x},t-\tau)
      \overline{Q}_I(0) n_\downarrow^{J^\prime}(\mathbf{y}, -\tau) \rangle \\
\label{eq:3ptspec}
&\quad= \sum_{\mathfrak{n},\mathfrak{m}} \sqrt{ Z_\mathfrak{n}^{J} Z_\mathfrak{m}^{J^\prime}}
    e^{-E_\mathfrak{n}(t-\tau)} e^{-E_\mathfrak{m} \tau}
\mbraket{\overline{\mathfrak{n}},\uparrow}{Q_I}{\mathfrak{m},\uparrow}\,,
\end{align}
where $\ket{\mathfrak{m},\uparrow}$ ($\ket{\overline{\mathfrak{n}},\uparrow}$) denote the spin-up
(anti)neutron states, $(\tau,t)$ are the Euclidean time intervals from the source to the operator
and the sink, respectively.

\begin{figure}[t]
  \centering
  \includegraphics[width=.4\textwidth]{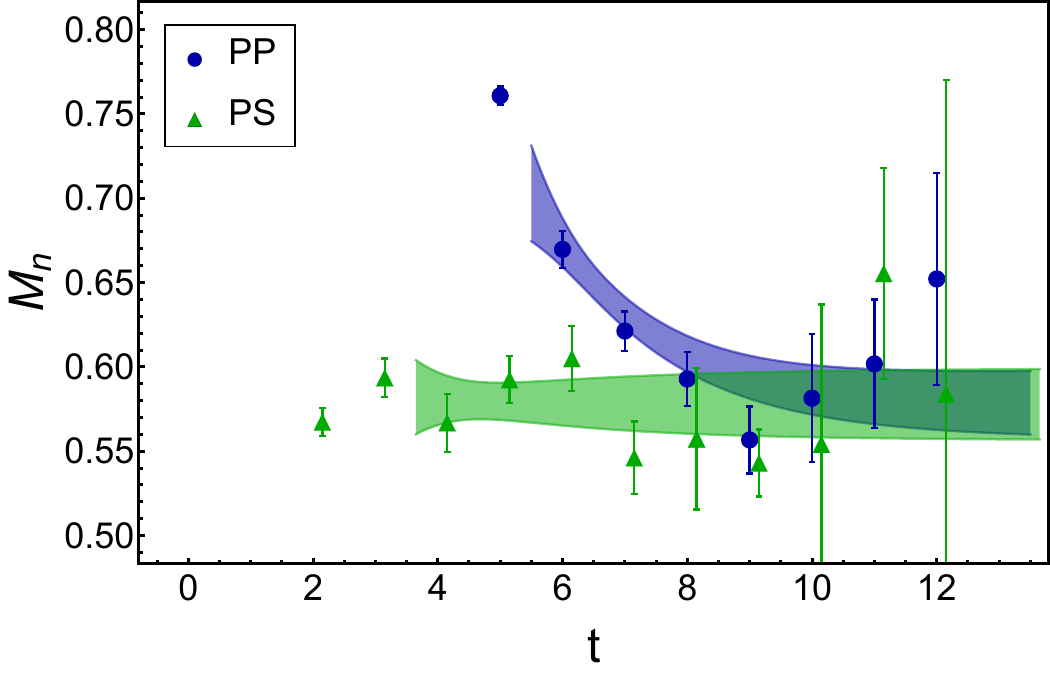}
  \caption{Neutron effective mass $M_n^{PJ}(t) = \ln\frac{G_{2pt}^{PJ}(t)}{G_{2pt}^{PJ}(t+a)}$
    determined from point-point and point-smeared ($J=P,S$) two-point correlation functions.
    Lattice data points are shifted for visibility and compared to
    two-state fits (shaded bands).
    The asymptotic $t$ result is compatible with the nucleon mass
    when converted to physical units, indicating negligible discretization and finite volume effects.
    \label{fig:2pt_meff}}
\end{figure}

\begin{figure}[t!]
  \centering
  \includegraphics[width=.4\textwidth]{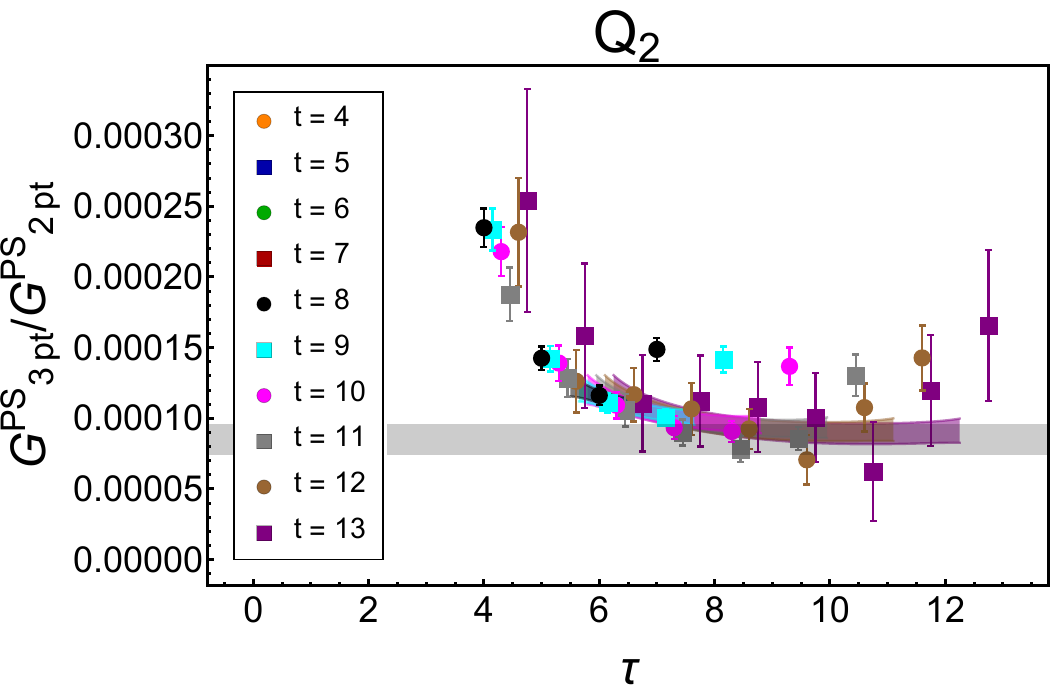}
  \includegraphics[width=.4\textwidth]{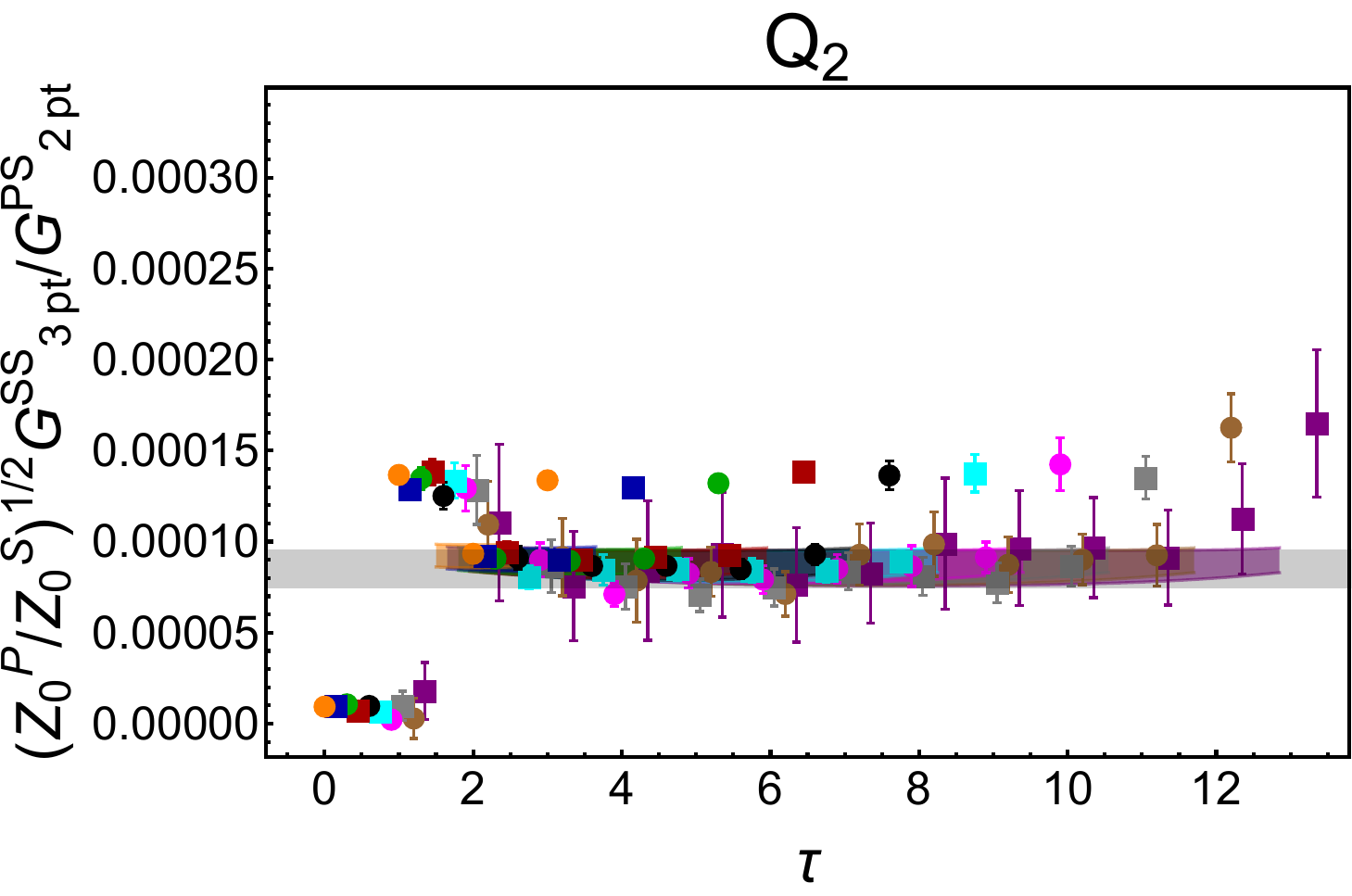}
  \caption{Ratios of three-point correlation functions for operator $Q_2$ to two-point functions
    \emph{vs.} operator insertion time $\tau$ and 10 different source/sink separations $t$.
    Lattice data points are shifted for visibility.
    These data are compared to two-state fits (colored shaded bands) used to extract the ground state
    bare matrix elements, which are shown with statistical uncertainty for a particular value of fit
    ranges (horizontal bands).
  \label{fig:3pt_Q2_ratio}}
\end{figure}

The $\nnbar$ ground-state matrix elements are extracted with two-state fits using
optimal shrinkage~\cite{LEDOIT2004365}, variable projection (VarPro)~\cite{varpro0,*varpro1}, and
weighted averages of results for a variety of source/sink separations that are
described at length in our companion paper~\cite{Rinaldi:2019thf}.
First, $G_{2pt}^{PP}$ and $G_{2pt}^{PS}$ are fitted in order to determine $E_{0,1}$ and $\sqrt{Z_{0}^{P,S}}$.
An example fit is shown in Fig.~\ref{fig:2pt_meff}.
These results are subsequently used to extract matrix element results from linear fits to $G_{3pt}^{SS}$ and $G_{3pt}^{PS} + G_{3pt}^{SP}$.
Ten different source/sink separations are included in order to isolate and remove excited state effects.
Ratios of $G_{3pt}^{JJ^\prime}$ to $G_{2pt}^{JJ^\prime}$ for operator $Q_2$,
that reach a plateau when the ground state saturates the correlation functions and excited state contributions have become negligible,
are shown in Fig.~\ref{fig:3pt_Q2_ratio} including fit results and statistical uncertainties.

In regularization schemes which do not violate chiral symmetry, operator mixing between
$Q_I^{(P)}$~(\ref{eqn:op_nnbar_123}--\ref{eqn:op_nnbar_567}) is forbidden, as in the continuum
perturbation theory calculation of Ref.~\cite{Buchoff2016}.
Since quark mass, residual chiral symmetry breaking, and nonperturbative effects might lead to
operator mixing on a lattice, we compute the renormalization and mixing of these operators
nonperturbatively using the regularization-invariant-momentum (RI-MOM) scheme~\cite{Martinelli:1994ty}.
RI-MOM renormalization factors $Z_{IJ}^{\text{RI}}$,
$Q_{I}^{\text{RI}}=Z^{\text{RI}}_{IJ} Q_J^{\text{bare}}$ at momentum $p$ are defined as
\begin{equation}
\label{eq:RIMOMdef}
\big[Z_q^{\text{RI}}(p)\big]^{-3} \, Z_{IJ}^{\text{RI}}(p) \, \Lambda_{JK}(p) = \delta_{IK},
\end{equation}
where $Z_q^{\text{RI}}$ is the quark field renormalization and $\Lambda_{JK}(p)$ are
amputated Green's functions of the lattice operators $Q_J$ and quark fields carrying momenta $\pm p$
projected onto the spin-color-flavor structure of $Q_K$.
All steps to calculate the renormalization factors numerically can be found in the companion
paper~\cite{Rinaldi:2019thf}.
We find that the matrices $\Lambda_{IJ}$ and therefore $Z_{IJ}^{\text{RI}}$ are diagonal in the
chiral basis (\ref{eqn:op_nnbar_123}--\ref{eqn:op_nnbar_567}) up to
$O(10^{-3})$~\cite{Rinaldi:2019thf}, thanks to chiral symmetry of the lattice action we use.
We neglect this residual mixing and identify $Z_{I}=Z_{II}$ below.

\begin{figure}[t!]
\centering
\includegraphics[width=.5\textwidth]{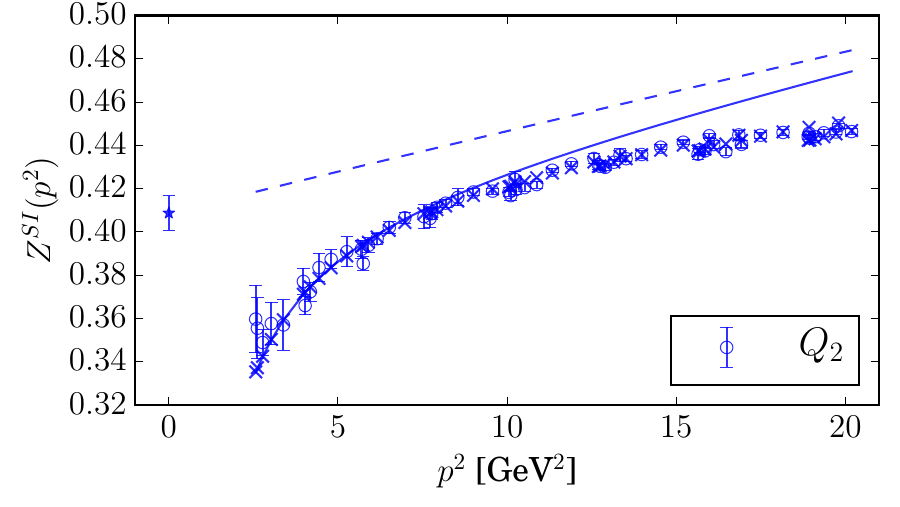}\\
\caption{RI-MOM ``scale-independent'' renormalization factors $Z^{\text{SI}}$ for the operator $Q_2$.
  Lattice data (circles) are fit to a constant (star) plus lattice artifacts:
  including $\propto(ap)^2$ (dashed line), $\propto(ap)^{-2}$ (solid line),
  and $O(4)$-breaking (crosses) terms~\cite{Blossier:2014kta}.
  \label{fig:nnbar_renorm}}
\end{figure}

The ``scale-independent'' combinations
\begin{equation}
  \begin{split}
    Z^{\text{SI}}_{I}(\mu_0,p) &= Z^{\text{RI}}_{I}(p)
    \bigg[ \frac{Z^{\text{RI}}_{I} (\mu_0)}{Z^{\text{RI}}_{I}(|p|)} \bigg]^\text{pert}
  \end{split}\label{eq:ZSIdef}
\end{equation}
of perturbative and nonperturbative factors have residual dependence on the lattice momentum $p_\nu$
due to nonperturbative effects $\propto p^{-2}$, discretization effects $\propto(ap)^2$, and
rotational symmetry breaking.
We analyze these effects following Ref.~\cite{Blossier:2014kta}, with a representative fit for
operator $Q_2$ and $\mu_0=2\text{ GeV}$ shown in Fig.~\ref{fig:nnbar_renorm}.
Fits with varying momentum ranges up to $1.6\le p\le4.5\,\text{GeV}$ are used to define central
values and stochastic and systematic uncertainties for $Z_{I}^{\text{RI}}$.
We have found no substantial difference between fits using 1-loop and 2-loop perturbative factors in
Eq.~\eqref{eq:ZSIdef}.
Fit details and similar plots for other operators can be found in Ref.~\cite{Rinaldi:2019thf}.
RI-MOM matrix element results are then converted to the $\msbar$(2 GeV) and $\msbar$(700 TeV) scheme
using 1-loop matching~\cite{Buchoff2016},
\begin{equation}
  \begin{split}
    \mbraket{\overline{n}}{Q_I^{\msbar}}{n}
      = \bigg[ \frac{Z^{\msbar}_{I,N_f=4}}{Z^{\text{RI}}_{I,N_f=3}}\bigg]^\text{pert}
        Z_{I,N_f=3}^{\text{RI}} \mbraket{\overline{n}}{Q_I}{n},
      \end{split}\label{eq:renorm}
\end{equation}
where the difference between $N_f=3$ and $N_f=4$ QCD is taken into account by matching
$\alpha_S$\cite{Bethke:2009jm} and operator normalization at the charm quark threshold $\mu = M_c$.
Statistical and systematic uncertainties of the renormalization factors and the bare matrix
elements are added in quadrature.
Higher-order matching uncertainties are estimated to be $\lesssim 7\%$ based on the size of 1-loop
matching effects \cite{Buchoff2016} and are neglected.
Final results for the $\msbar$ matrix elements are shown in Tab.~\ref{tab:renorm_me}
and compared to previous ``MIT bag model'' results.

\textbf{\textit{Phenomenological implications --- }}\label{sec:pheno}
In BSM theories where $\Delta B = 2$ transitions are permitted, experimentally observable $\nnbar$
oscillations are low-energy phenomena that can be described in an EFT containing only SM fields.
The low-energy EFT will include $\Delta B = 2$ terms involving the
operators $Q_I$ above,
\begin{equation}
  \begin{split}
    \mathcal{L}_{\nnbar} = \sum_{I=1}^7   \left(C_I(\mu)  Q_I(\mu) + C_I^P(\mu)  Q_I^P(\mu) \right),
  \end{split}\label{eq:Lnnbar}
\end{equation}
where the $C_I$ are numerical coefficients with mass dimension $(-5)$ that are predicted to
be non-zero in some BSM theories.
The $SU(2)_L$-singlet operators are EW-symmetric and their coefficients should scale as
$C_{1,2,3,4} \sim \lbsm^{-5}$ in naive dimensional analysis.
In contrast, the $SU(2)_L$ non-singlet operators $Q_{1,\ldots,7}^P$ and $Q_{5,6,7}^{(P)}$ can only
appear in an SM gauge-invariant Lagrangian in products with additional SM Higgs (or BSM) fields to
make them $SU(2)_L$ singlets.
Assuming the former, their coefficients should scale as
$C_{1,\ldots,7}^P(\lbsm) \sim v^2 \lbsm^{-7}$ and $C_{5,6,7} \sim v^4 \lbsm^{-9}$,
where $v$ is the vacuum expectation value of the Higgs field.
For $v \ll \lbsm$, this provides a significant additional suppression on $\nnbar$ oscillation rate
contributions from $SU(2)_L$-non-singlet operators.

The $\nnbar$ oscillation rate is given by the matrix element of the associated Hamiltonian between
neutron and antineutron states, which in the isospin limit of QCD simplifies to
\begin{equation}
\label{eq:taudef}
\tau_{\nnbar}^{-1}
  = \Big| \sum_{I=1,2,3,5} \widehat C_I(\mu) \mbraket{\bar{n}}{Q_I(\mu)}{n}\Big|\,,
\end{equation}
where $\widehat C_I = C_I - C_I^P$ for $I~=~1,\ldots,4$ and
$\widehat C_5 = (C_5 - C_5^P) + (C_6 - C_6^P) - \frac{2}{3}(C_7 - C_7^P)$.
Contributions involving $C_4^{(P)}$ vanish exactly in the isospin limit considered here, although in
principle isospin-violating $C_4^{(P)}$ contributions could play a role in particular BSM models.
Using the LQCD results above, the $\nnbar$ oscillation rate is given by
\begin{align}
\nonumber
&\tau_{\nnbar}^{-1} = \frac{ 10^{-9} \text{ s}^{-1} }{(700 \text{ TeV})^{-5}} \big|
    4.2(1.1) \widehat C_1^{\msbar}(\mu) - 8.6(1.5)  \widehat C_2^{\msbar}(\mu) \\
\label{eq:taufid}
  &\quad + 4.5(1.1) \widehat C_3^{\msbar}(\mu) + 0.096(43) \widehat C_5^{\msbar}(\mu)
  \big|_{\mu = 2\text{ GeV}}\,.
\end{align}
It is noteworthy that, in addition to the $(v / \lbsm)^2$ suppression,
contributions from $\widehat C_5$ are suppressed by almost two orders of magnitude compared to
those from $\widehat C_{1,2,3}$ due to the relative sizes of the associated QCD matrix elements
computed here.

Predictions for non-zero $\tau_{\nnbar}^{-1}$ arise in some BSM theories explaining the
matter-antimatter asymmetry of the universe and other outstanding problems of the SM and cosmology.
An example is provided by $\nnbar$ oscillations in left-right symmetric gauge theories where the SM
gauge group is embedded in $SU(2)_L\times SU(2)_R\times SU(4)_C$ with $(B-L)$ acting as a fourth
color~\cite{Mohapatra:1980qe,Babu:2008rq,Babu:2013yca}.
Post-sphaleron baryogenesis occurs after a colored BSM scalar field develops a $(B+L)$-breaking
vacuum expectation value that leads to Majorana neutrino masses and $\nnbar$
oscillations.
The $\nnbar$ oscillation rate in this model only involves $Q_1$ at tree-level and
$\tau_{\nnbar}^{-1}$ is given by the first term in Eq.~\eqref{eq:taufid}.
The reach of current and future $\nnbar$ oscillation experiments into the parameter space of this
$SU(2)_L\times SU(2)_R\times SU(4)_C$ model is 4 to 5 times higher according to LQCD than the bag model.
In turn, since oscillation probabilities are $\propto \tau_{\nnbar}^{-2}$~\cite{Phillips2016},
this model would predict 16 to 25 times larger number of observed events in quasi-free neutron experiments.
Updated phenomenological studies of this and other BSM theories of low-scale baryogenesis are
needed to determine the reach of current and future $\nnbar$ experiments into BSM parameter space.

\textbf{\textit{Conclusions --- }}\label{sec:conclusions}
An LQCD calculation of six-quark matrix elements is presented that provides, for the first time,
renormalized $\nnbar$ transition matrix elements in the $\msbar$ scheme at 2 GeV.
These renormalized results with well-defined scale dependence are required to reliably connect
experimental measurements of $\tau_{\nnbar}$ to the baryon number-violating new physics scale
\lbsm.

This calculation is performed with physical pion masses, a chirally-symmetric fermion
discretization, and a large spacetime volume.
Ground-state matrix elements are extracted using
2-state fits and systematic uncertainties associated with excited states are estimated
through variation of the fitting region.
Finite volume effects are predicted by chiral EFT to be $\lesssim 1\%$~\cite{Bijnens2017} for the
$L\approx5.45\text{ fm}$ volume used in this study.
Even though we only use one lattice spacing $a \approx 0.114$ fm, discretization effects are
expected to be small because of automatic $\mathcal{O}(a)$ improvement due to the chiral symmetry of
the DWF action.
Additional systematic effects due to the uncertainty in the lattice scale $\delta a / a\approx1.3\%$
are negligible.
The error budget of our final results is dominated by limited statistics.
Statistical uncertainties and systematic uncertainties associated with discretization and
finite volume effects can be improved in the future using additional ensembles~\cite{Blum:2014tka}.
However, they are very unlikely to change the dramatic impact that QCD effects have on these
six-quark matrix elements.

To conclude, there has been recent phenomenological interest in $\nnbar$ oscillations and the
possibility of new searches for $\nnbar$ vacuum oscillations and transitions in nuclei at ESS, DUNE,
and other experiments~\cite{Milstead2015,Frost2016,Hewes2017,Fomin2018}.
The magnitudes of electroweak-singlet $\nnbar$ transition matrix elements are 4-8 times larger than
those computed in the ``MIT bag model''~\cite{Rao1982}.
Experiments should consequently observe 16-64 times more neutron-antineutron oscillation events
for fixed BSM physics parameters.
Future searches for $\nnbar$ oscillations will be able to probe the parameter space of several viable baryogenesis
scenarios~\cite{Grojean2018}.
Our results, despite being pioneering, indicate that experimental searches of $\nnbar$ transitions
are about 1 to 2 orders of magnitude more sensitive to baryon
number violating interactions in BSM physics than previously expected and will be able to put more
stringent constraints on various baryogenesis mechanisms.




\textbf{\textit{Acknowledgements --- }}
We are indebted to Norman Christ, Bob Mawhinney, Taku Izubuchi, Oliver Witzel, and the rest of the
RBC/UKQCD collaboration for access to the physical point, domain-wall lattices and propagators used
in this work.
We would like to thank Yuri Kamyshkov, Rabi Mohapatra, Martin Savage, Steve Sharpe, Robert Shrock,
Mike Snow, Brian Tiburzi, and hosts of others for discussions over the duration of this project.
Computing support for this work was provided in part from the LLNL Institutional Computing Grand
Challenge program and from the USQCD Collaboration, which is funded by the Office of Science of the
US Department of Energy.
This research used resources of the Argonne Leadership Computing Facility, which is a DOE Office of Science User Facility supported under Contract DE-AC02-06CH11357.
This work has been supported by the U.~S.~Department of Energy under Grant Nos. Contracts
DE-AC52-07NA27344 (LLNL), DE-FG02-00ER41132 (INT).
Brookhaven National Laboratory is supported by the U.~S.~Department of Energy under contract
DE-SC0012704.
SS acknowledges support by the RHIC Physics Fellow Program of the RIKEN BNL Research Center.
ER is supported by the RIKEN Special Postdoctoral Researcher fellowship.
MLW was supported by a MIT Pappalardo Fellowship and acknowledges support by the U.S. Department of
Energy, Office of Science, Office of Nuclear Physics under grant Contract Number DE-SC0011090.



\bibliography{NNbar}

\end{document}